# PARKINSON'S DISEASE PATIENT REHABILITATION USING GAMING PLATFORMS: LESSONS LEARNT

Ioannis Pachoulakis[1], Nikolaos Papadopoulos[2] and Cleanthe Spanaki[3]

[1,2]Department of Informatics Engineering, Technological Educational Institute of Crete, Heraklion, Crete, Greece
[3]Department of Neurology, Faculty of Medicine, University of Crete, Heraklion, Crete, Greece

### ABSTRACT

*Parkinson's disease (PD) is a progressive neurodegenerative movement disorder where motor dysfunction gradually increases as the disease progress. In addition to administering dopaminergic PD-specific drugs, attending neurologists strongly recommend regular exercise combined with physiotherapy. However, because of the long-term nature of the disease, patients following traditional rehabilitation programs may get bored, lose interest and eventually drop out as a direct result of the repeatability and predictability of the prescribed exercises. Technology supported opportunities to liven up a daily exercise schedule have appeared in the form of character-based, virtual reality games which promote physical training in a non-linear and looser fashion and provide an experience that varies from one game loop the next. Such "exergames", a word that results from the amalgamation of the words "exercise" and "game" challenge patients into performing movements of varying complexity in a playful and immersive virtual environment. Today's game consoles such as Nintendo's Wii, Sony PlayStation Eye and Microsoft's Kinect sensor present new opportunities to infuse motivation and variety to an otherwise mundane physiotherapy routine. In this paper we present some of these approaches, discuss their suitability for these PD patients, mainly on the basis of demands made on balance, agility and gesture precision, and present design principles that exergame platforms must comply with in order to be suitable for PD patients.*

## 1. INTRODUCTION

Parkinson's disease (PD) is a neurodegenerative condition which affects parts of the brain that control body movement. More precisely, PD results from the loss of dopamine-producing neurons in a mid-brain region known as "substantia nigra", responsible among others for the smooth and purposeful coordination of body muscles. It is still unclear why these cells are lost, making PD so far untreatable [1], [2]. The ailment was named after James Parkinson, who first recorded and reported the symptoms. According to the European Parkinson's Disease Association (EDPA) [1], around 6.3 million people suffer from the disease worldwide. Symptoms progress slowly but irreversibly, so that late stages of the disease are more abundant in the elderly (>60 year old) population, while only approximately 10% of PD patients are under 50 years old. In early stages, PD commonly affects motor function, whereas cognitive, behavioral and mental-related symptoms are usually met at more advanced stages [3]. Non motor-related symptoms include sleep disturbances, depression, anxiety, psychosis and visual hallucinations, cognitive impairment, pain and fatigue. In addition, the four distinct, fundamental motor symptoms of the disease can be grouped under the acronym TRAP: Tremor, Rigidity, Akinesia (or bradykinesia) and Postural instability [4]. These symptoms often develop in different combinations and result in hindering common daily activities, troubling patients' social relationships and reducing the quality of life, especially as the disease progresses [5], [6].

Tremor can be distinguished in tremor at rest and postural tremor. Tremor at rest is the most ordinary symptom, appearing at some point during the disease among 75% of the patients population [7], making it the most distinctive and easily recognized sign of the disease. It is described [1] as an unintentional and rhythmic movement of body parts (limps, head and face sections) while relaxed. Meanwhile, postural tremor is associated with body movement (e.g. walking) or controlled still posture (e.g., standing) and can easily be misdiagnosed as essential tremor in the absence of other symptoms [8]. Rigidity, on the other hand, describes the inability of limb muscles to relax, providing high resistance during passive movement of a limb, also known as cogwheel phenomenon. In more advanced stages, the muscles of those limbs can be described as stiff and highly inflexible. PD patients may also suffer from pain due to the rigidity, such as 'painful shoulder', one of the trait characteristics for PD [4]. According to EPDA [1], akinesia (or bradykinesia) is one of the three main traits of PD and refers to the slowness of PD patients in carrying out a voluntary movement, rather than initiating one. Akinesia affects 78-98% of the PD patient population [8], troubling them during the entire course of the disease. To detect akinesia, PD patients are usually asked to repeatedly perform rapid movements. Finally, postural instability expresses the lack of postural reflexes during standing, walking or interacting with objects in space [9] and is usually absent in early stages of PD. This last symptom depends on the severity and course of the disease [10] and is highly correlated to the frequency of patients' falls. Postural instability combined with akinesia can be an especially dangerous combination which can lead to severe injuries.

To assess the progress of PD as well as a given patient's level of disability, several rating scales are employed. In 1967 Hoehn and Yahr [11] proposed a simple and easy to apply rating scale which recognizes three practical classification types for Parkinsonism (primary, secondary and indeterminate). The Hoehn and Yahr scale utilizes a 5-stage system for grading the severity of the condition which ranges from insignificant motor impairments in the first stage, to severe impairments in late stages. The scale focuses on observations such as unilateral or bilateral expression of the disease as well as the degree of postural reflex impairment. Several additional stages have been proposed to add detail and granularity to the initial scale. These stages include non-motor aspects of the disease and describe motor aspects more precisely [12]. In 1980 the Unified Parkinson's disease Rating Scale (UPDRS) [13] combined several elements from previous scales and has been further updated by the Movement Disorder Society (MDS) [14] to include new aspects of non-motor symptoms. The UPDRS scale consists of three major sections which evaluate significant areas of disability (Part I: Mentation, Behavior and Mood; Part II: Activities of daily living; Part III: Motor Function). The scale is accompanied by a fourth section that evaluates complications in treatment. UPDRS is the most widely used clinical rating scale and according to EPDA is commonly used in tandem with the Hoehn and Yahr scale as well as the Schwab and England Activities of Daily Living (ADL) scale. It must be noted that the ADL scale provides a useful measure of a person's capability in performing daily activities, and as a result of that person's independence.

Although a cure has not yet been discovered for PD, medications usually help control the symptoms and maintain body functionality at reasonable levels through a patient's lifetime. According to information provided by the American Parkinson Disease Association (APDA) [15] six categories of drugs are proposed for PD condition therapy: levodopa, dopamine agonists, MAO-B inhibitors, COMT inhibitor, anticholinergic agents and amantadine. Standard medication paths cannot be easily achieved, as the course and symptoms of the disease varies significantly among patients [16]. Based on current clinical practice, levodopa is considered as the most efficient drug for improving PD-related motor symptoms, although large doses over an extended period of time have been connected to the development of involuntary abnormal movements called dyskinesias that further aggravate walking ability and motor function. Recent common practice starts patients on agonists, postponing levodopa for later stages when motor symptoms are not satisfactorily controlled [17]. In more advanced stages, combinations of levodopa, dopamine agonists, COMT inhibitors and MAO-B may be used to control symptoms and achieve optimal results [16].

## 2. BENEFITS OF PHYSIOTHERAPY FOR PD PATIENTS

Adding to the value of medical treatment, physiotherapy appears highly effective in controlling PD-related symptoms. A number of clinical facilities and associations focusing on PD provide physical activity guidelines, suggesting daily activities and tasks, even diet schedules. For example the Parkinson Society of Canada [18] provides online detailed instructions on how to correctly perform stretching and other physical exercises. Exercise interventions in randomized controlled trials [19] show that physical exercise such as stretching, aerobics, unweighted or weighed treadmill and strength training improves motor functionality (leg stretching, muscle strength, balance and walking) and quality of life. Interestingly, one training program [20] was conducted in patients' homes instead of at a clinical facility using exercises tailored to the condition of each patient. In addition, a physiotherapist was visiting patients on a weekly base. Participants kept records of fall events for the duration of the program. Analysis of these records reveals lower fall rates for patients that follow home-based exercise programs with detailed preparation and documentation compared to those who do not. These results are corroborated by a different study [21] which employed experimental balance training including self-destabilization exercises, externally-forced destabilization exercises and coordination of leg and hands during walking. These exercises improved postural stability and boosted patients' confidence as a result of the reduced frequency of falls. In fact, benefits in postural stability as a result of that exercise program were maintained for at least one month after conclusion of the program.

Significant fringe benefits resulting from home-based rehabilitation include significant cuts in treatment costs. Indeed, PD patients must frequently attend physiotherapy sessions to either notice an improvement or maintain the gains from clinical rehabilitation programs [22]. In fact, Calgar, et al. [23] point out the effectiveness of home-based, structured physical therapy exercise programs tailored to the individual patient, as witnessed by measurable improvements in motor capability. In addition, the "training BIG" strategy for PD rehabilitation has also shown promising results. For example, exercises that focus on amplitude training [24] can lead to faster upper and lower limb movements – in this case, participants repeatedly performed various exercises using maximum range of motion (maximum amplitude). The exercise sets employ the entire body of the patient, both in seated and standing posture and include BIG stretches (i.e. reach and twist to side) and reparative BIG multidirectional movements (i.e. step and reach forward).

Several other training programs have been applied to the PD condition. Prominent among those is Tai Chi, a form of martial art based on gaining balance through the controlled and continuous movement of the body's center of mass. As several studies have shown the benefits of Tai Chi in strength, balance, and physical function in healthy older adults, Li et al. [25] conducted a study to evaluate Tai Chi's potential in improving postural stability for PD patients. The randomized trials included three independent patient groups: a Tai Chi group, a resistance training group, and a body stretching group. Tai Chi participants showed more significant improvements both in balance and in maximum excursion compared to patients of the remaining two groups. In addition, Zhou, et al. in their systematic literature review [26] conclude that Tai Chi seems to significantly improve motor and balance impairments for PD patients, although larger-scale samples and high-quality randomized control trials are necessary to make this statement conclusive. Another promising rehabilitation exercise trial was conducted by Combs et al. [27] using box training. While the number of patient participating in this trial was relatively small (six patients), the patients showed both short and long term improvements in balance, gait, daily activities and quality of life, although more advanced stage participants seem to require more persistence and time to reap these benefits.

## 3. TECHNOLOGY-SUPPORTED PHYSIOTHERAPY

PD patients attending long-term rehabilitation exercise programs tend to get bored of the same daily physical routine [28]. In fact, Mendez et al. [29] propose to investigate the learning potential of PD patients using new therapeutic strategies and validating their utility. Indeed, it is worth examining the potential benefits reaped by PD patients through their participation to exergames - computer games indented to be used as an exercise tool, using motion capture systems like Sony Playstation Eye, Nintendo Wii and Microsoft's Kinect. Such games use audio and visual cues in loose, virtual reality (VR) environments and offer an enormous motivating potential because they do away with repeatability (the seed of boredom) and engage users in immersive, goal-oriented scenarios. Motion capture devices provide an interface to the virtual world and can be programmed and tuned to provide real-time information about the specific interactions.

VR interaction is promising in terms of rehabilitation not only for people with the PD condition, but also for people with motor degenerative conditions. A resent review [30] conducted by Vieira et al. evaluated various studies in the literature with an eye on the possible benefits of VR-based systems for PD patients. It concluded that VR can not only be used as a therapeutic tool, but can also play a significant role in controlling and regaining motor function, mobility and cognitive capacities as well as balance. In addition, participants in all home-based trials showed improvements in all post-training tests, supporting the idea of using VR as a therapeutic home-based tool. The authors continue to discuss the benefits of VR-based games on physical neuro-rehabilitation supporting that visual and auditory cues may stimulate a player's reaction at a cognitive level. Indeed, brain re-training seems possible at some level, in the sense of rerouting brain activity through alternative neuron paths. Other recent studies seem to corroborate not only the feasibility but also the benefits of exergames for PD patients, as a highly effective rehabilitation practice [31]. At the same time, extreme care and forethought must characterize the design of exergames specifically for PD patients. For example, exergames must provide motivation and positive feedback, be progressively more challenging but also adaptable to a specific patient's condition and all that within a "maximal safety net" from the design stage and not as an afterthought so as to minimize and even eliminate accidents such as falls.

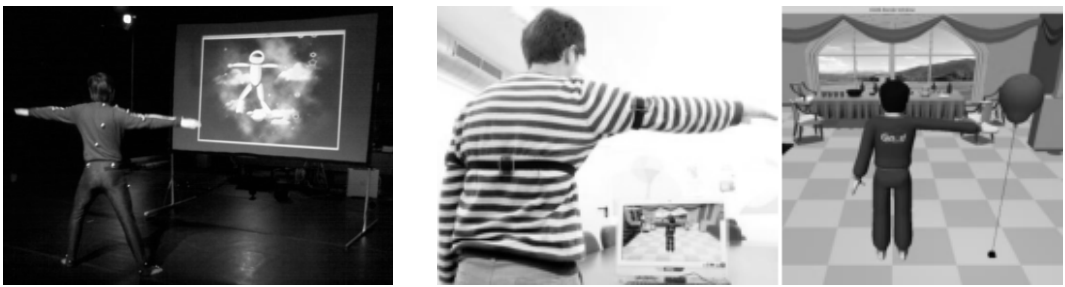

Figure 1: Left – screen shot from the sky scenario of Yu et al, where the avatar stands on a cloud to reach falling stars. Right – a user wearing 3 IMUs (torso and upper left/right arm) plays the Touch'n'Explode game (adopted from Tous et al)

Yu et al [32] developed a real-time multimedia environment aiming at PD patient rehabilitation. The system was developed using a ten near-IR camera Motion Analysis System to capture a patient's body in 3D. To successfully capture movement, retro-reflected markers were attached to the patient's body (left panel in Figure 1). Guided by visual and auditory cueing, patients were called to execute several exercises based on the BIG strategy [24]. At the same time, incoming data streams were analyzed in real-time via motion analysis software and were mapped onto an on-screen avatar. However, the proposed solution is not cost effective due to the need of installing multiple infrared cameras and calibrating the system. In addition, it is hard for PD patients to use the system on their own without external help, due to the fact that the markers must be attached to their body at specific points every time they need to exercise.

In the context of the EU-funded CuPiD project, Tous et al. extended the tele-rehabilitation platform Play For Health (P4H) to create exergames for PD patients [33]. Three games were created: Touch 'n' Explode (pop objects in the virtual environment), Stepping Tiles (step on the tiles of the kitchen) and Up 'n' Down (sit-stand exercise). Motion capture used a body-area network (BAN) of wearable sensors developed specifically for the project along with the necessary algorithms for posture recognition. Patient movements were translated to 3D avatar movements in a virtual environment (right panel in Figure 1). The system was also able to detect freezing motion incidents.

Another line of efforts evaluated the Nintendo Wii gaming system as a possible rehabilitation tool for PD ([31] presents a systematic review). The Wii system uses a handheld controller to communicate with a console, usually installed by the display monitor / TV in-front of the user. Data such as rotation and acceleration of the controller is sent to the console wirelessly. In addition, the "balance board", a rather common extension component used by several Wii games contains several sensors which measure the mass of the player and his/her center of gravity. Case studies evaluated Wii as an off-the-self, affordable solution that already provided major game releases in rehabilitation. Data related to gait, balance, cognitive reaction and functional mobility were recorded and presented. In most cases, however, the Wii Balance board extension had to be used, as it is required by the majority of Wii games.

Zettergren et al [34] evaluate three exergames (Penguin Slide, Table Tilt, Balance Bubble) and four activity games (Free Step, Island Cycling, Obstacle Course and Rhythm Parade) with the Nintendo Wii platform. Results show significant benefits on gait speed and timed up-and-go, as well as improvements on Berg's balance scale, without signs of (psychological) depression observed during the trial. An additional study that exhibited improvements among PD patients in terms of gait and balance was conducted by Mhatre et al [35]. The trial used Wii's balance board extension of the Wii gaming system for the Marble, Skiing and Bubble games. Esculier [36] et al used similar Wii games to evaluate the benefits in balance for PD patients and compared the results against a set of healthy elderly individuals. The trial showed improvements in most static and dynamic balance aspects for both user groups. Four out of five trial participants liked the games whereas the fifth participant was neutral on the matter.

Pompeu et al [37] investigated whether PD patients can improve their performance on the Wii gaming system and compared the effects of Wii-based motor and cognitive training with balance exercise therapy in the aspect of independent performance for daily life. Tested Wii games included balance games (Table Tilt, Tilt City, Penguin Slide and Soccer Heading) as well as static balance games (Torso Twist and Single Leg extension), while stationary gait was practiced in games like Rhythm Parade, Obstacle Course, Basic Step and Basic Run. Results showed that the PD patients improved their performance in each category of Wii-based games. Although Wii-based motor and cognition training had good impact on independent performance of activities for daily life, the same results were observed with the balance exercise therapy executed in a real environment.

Hertz et al [38] on the other hand, used the Wii games Tennis, Boxing and Bowling to determine the effectiveness of the specific gaming system on both motor and non-motor symptoms for PD patients. The games were chosen as out-of-the-box solutions bundled with the Wii system (lower cost than buying later as add-ons) with familiar movements to the subjects of the study. Participants witnessed improved motor and non-motor aspects as well as improvements in quality of life.

Most aforementioned studies using commercial Wii games demonstrated improvements in gait, motion and balance among PD patients. In addition, a number of studies find that benefits reaped during program participation were maintained for a time period after the program's completion. It must be mentioned, however, that several games developed for the Wii platform require the balance board raised platform as an extension. For example, in Free Step patients step on and off the balance board to follow the game's sequential activities. It should be noted, however, that the balance board is a risky piece of hardware for PD patients, as it may lead to falls [31]. In addition, all Wii games require a handheld controller to capture the player's movement, which may be troublesome for at least some PD patients. Therefore, whereas commercial Wii exergames seem to offer low-cost rehabilitation opportunities, they may not be suitable for all PD patients. In fact, Mendes et al [29] record that PD patients have trouble in learning and retention using some Wii commercial games when compared with healthy individuals, resulting in poor performance in those games. In addition, study participants failed to improve their performance for games requiring fast reaction, which case subjects naturally lacked as a result of cognitive and/or motor problems inherent to the PD condition. It follows that game performance improvements for the PD population depends on the demands of the specific Wii game played.

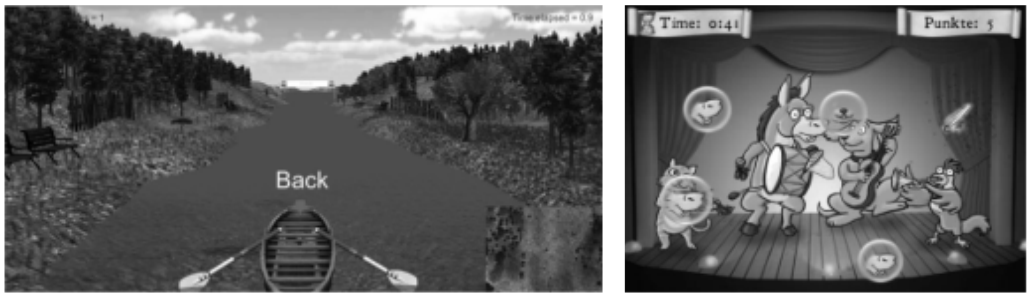

Figure 2: Left – Rowing game setup (adopted from Paraskevopoulos et al). Right – screenshot of the game Town Musicians (adopted from Assad et al)

Wii-based rehabilitation games tailored to PD patients were developed by Paraskevopoulos et al [39]. Two games intended to improve speed, rigidity, range of movement and bilateral mobility were developed based on PD-specific movements extracted from bibliography. In the first game the patient controls a two-paddle rowboat from a seated position to reach a specific point in a certain amount of time (left panel in Figure 2). In the second game (water valve mini golf game) the player rotates a virtual valve to release and guide a ball through a pipe into a hole. Data retrieved from the sensor's accelerometer and gyroscope were algorithmically combined to provide orientation and linear motion results so as to map the user's movements on screen actions.

Evaluating the advantages and disadvantages of similar (to the Wii) technologies, Assad et al [28] examined the Sony Playstation Eye system as a potential rehabilitation tool for PD patients. They developed WuppDi!, a collection of five PD-oriented motion-based games within a playful environment. Most of these games required one or two (one for each hand) markers in form of a glove or wooden stick to interact with virtual game objects. Participants welcomed this approach as a physical activity albeit having trouble handling the input devices. Earlier trials with the Wii

International Journal of Biomedical Engineering and Science (IJBES), Vol. 2, No. 4, October 2015gaming system had shown that PD patients weren't able to manipulate the handheld controller effectively, as it required moving the controller and pushing buttons on it at the same time.

The latter two studies revealed a necessity for developing exergames specifically aimed to PD rehabilitation. However, emphasis must be given to gaming systems that accurately capture full body motion without the need of external handheld or wearable devices. In their review, Gillian et al [31] propose that platforms such as the XBox Kinect which do not require raised platforms, handheld controllers and/or body markers need to be evaluated with respect to the rehabilitation opportunities with maximal safety for PD patients.

The Kinect sensor requires no external input controllers and can capture the motion of the entire human body in 3D, using an RGB camera and a depth sensor. Players can manipulate game interactions by moving their body in front of the sensor, i.e., the human body is the controller. Kinect-based exergames tailored to PD are however sparse due to the fact that the technology is still relatively new. The first generation of Kinect sensors was released in November of 2010 and was paired to the Xbox 360 console, while in June 2011 Microsoft released a Software Development Kit (SDK) to allow the creation of Kinect-based applications.

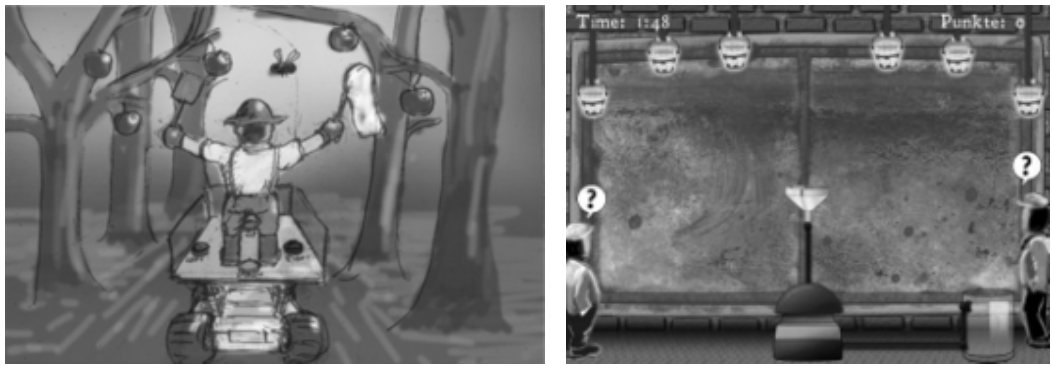

Figure 3: Left – Scene from the farmer game of Galna et al. Right – screenshot from the loose cooperation game of Hermann et al.

Galna et al [40] developed a Kinect game for PD patients which included specific upper and lower limb movements to improve dynamic postural control. The upper torso of the player is mapped onto a farmer avatar driving a tractor who collects fruits and avoids obstacles in a 3D environment (left panel in Figure 3). Fruit collecting is achieved by specific arm movements, choosing to move the right or the left hand depending on the color of the fruit. Large steps (front, back and sideways) with one foot centered guide the tractor to avoid obstacles in the form of sheep, high wires, birds, etc. To maintain patient motivation, the game has several levels of difficulty: game complexity increases from simple hand movements for low levels through more complex activities combining cognitive decisions and physical movements all the way to dual tasking (simultaneous hand and foot work). The game's design principles resulted from a workshop where participants played and evaluated a wide range of commercial games developed for Microsoft's Xbox Kinect and Nintendo's Wii. The workshop once more revealed the aforementioned difficulties of patients to interact with Wii's handheld controller and balance board. A pilot test led to several useful conclusions, extracted through interviews and questionnaires, in terms of gameplay design, feasibility and safety. The authors conclude that the Kinect sensor seems both safe and feasible for PD patients, although its use as a home-based rehabilitation solution needs further investigation.

Another multiplayer game also based on the Kinect sensor was developed by Hermann et al [41] to investigate whether gamer cooperation can lead to improvements in communication and coordination. Using hand movements, participants had to collect buckets of water in a flooded area to reveal an object underneath. The game was played in two different modes. In the first mode (loose cooperation) both players drained the area (right panel in Figure 3), while in the second mode (strong cooperation) one player would drain, while the second player was expected to reveal the hidden object. Gameplay was recorded to reveal information regarding the discussion and level of cooperation between the participating patients. Subsequent analysis showed that multiplayer games are feasible for this kind of target group and that asymmetric roles (strong cooperation) can motivate communication between participants and lead to a better game experience.

## 4. DISCUSSION - LESSONS LEARNT

To design effectively for PD patients requires that a number of good practices and design principles are followed, as a result of understanding the symptoms of the PD condition. These practices and principles are supported by the results of the studies presented in the previous sections and are presented directly below.

### 4.1. Gaming System Suitability – Safety First!

In their review Gillian et al [31] support that exergame solutions for PD patients should avoid raised platforms and handheld controllers such as those employed by the Nintendo Wii and Sony PlayStation Eye gaming systems. Because lack of postural stability characterizing the PD condition may lead to falls and severe injuries, Gillian et al suggest avoiding raised platforms (such as balance board) which present additional trip and fall risks. It follows that patient safety is maximally preserved with gaming systems which provide full body motion tracking without additional / external add-ons that users must interact with. Low cost solutions with minimal footprint that can easily be set up either in a clinical setting or a home environment would naturally be preferred. Still, all such systems must be evaluated in the terms of patient safety.

### 4.2. Patient-specific Game Solutions

A number of studies, e.g., [28], [29], [31], [39], [40] have showed that while off the shelf, low cost gaming systems can easily exist in both clinical and home environments, not all commercial exergames developed for those systems are suitable for the PD condition. For example, evaluation trials conducted in [28] and [40] showed that PD patients wearing gloves or using handheld controllers with buttons encountered difficulties interacting with the virtual game environment. In addition, Mendes et al [29] deduce that PD patient performance in commercial exergames depends on the motor and cognitive requirements of the game itself. To further elaborate on this observation, PD participants failed to improve game performance compared to a control group of elderly healthy individuals in three of the commercial Wii games that were tested. It is therefore advisable that exergames include specifically designed movements drawn from existing recommended PD training programs, an approach followed, for example, in [39]. To further improve motor function (range of motion, balance, postural reflex and strength) the training BIG strategy [24] supports big, purposeful and smooth movements to increase the speed of upper and lower limbs, similar to the philosophy of Tai Chi (e.g., [25] and [26]). The conclusions of the preliminary trials of Assad et al [28] have also showed that PD patients may be frustrated by tremor that hinders performing slow and accurate arm movements.

## 4.3. Personalization and Adaptability

Motor skills and especially the range of extent for limbs vary significantly among PD patients and depend on several factors such as the course of the disease, medication and everyday physical activity. Accordingly, exergame movements must be scaled and calibrated based on the individual user's motor skill set [28]. In addition, exercise adaptations may vary from more challenging strict mappings to looser ones so as to avoid patient fatigue or frustration e.g., by repetitive failure of a task [32]. Therefore, not only captured movements must be adaptable, but also the difficulty of specific game tasks, such as game speed [40].

## 4.4. Game Scenario Variety

In the introductory section of the present work it has been mentioned that PD is more prevalent to an aging population section (> 60 years old). Most of these elderly people may not have experienced the computer revolution first hand and have little or no experience in gaming systems. To maintain interest for this significant population of PD patients, games must be simple in terms of design and scenario and must contain familiar concepts [28]. Indeed, Galna et al [42] found that patients were not attracted by complex adventure-type or science fiction game scenarios, but instead were drawn to games based on real-life experiences that used cartoon-style graphics, which they found more familiar and appealing. In addition, a number of patients showed preference to outdoor activity scenarios [40]. Game simplicity was also preferred in the sense that game assets are easily and immediately recognizable during gameplay and their orientation inside the game scene is made obvious so as not to distract from the goal of improving motor performance.

Exergames must also contain clear instructions and challenging goals to maintain motivation to improve motor performance. During the workshop carried out in [40] patients showed overwhelming preference to games that aren't too complex or too fast in reaction speed, while one patient mentioned that the game was too easy. A way to allow for a wide gamut of user preferences and abilities is to build multiple game levels. Less demanding low levels introduce simpler / slower tasks intended to familiarize with the game and/or avoid frustration, whereas more advanced levels may demand increased complexity, multitasking and faster reaction. Such is the approach of e.g., Assad et al [28], whose games show different levels of complexity, from really simple ones which focus solely on body movement to more difficult levels which require coordination and concentration.

One can thus infer that challenging game scenarios will stimulate both motor functionality (balance, strength and postural reflex) and cognition (e.g., making a correct, timely decision). Following suggestions of physiotherapists that participated in their system design, Yu et al [32] employed the concepts of accuracy and timing to contain and reduce symptoms related to bradykinesia. Patients had to follow and execute instructive movements in a specific amount of time. Time threshold values were chosen carefully to motivate patients to react faster but were also sufficiently tolerant and flexible to accommodate individual player capabilities.

## 4.5. Visual and Auditory Feedback

Positive feedback, in the terms of visual and auditory effects, is very important to all games and even more so to exergames tailored to PD patients. Visual and auditory cues accompanying score bumps or drops, communicating the points gained or rewarding for the completion of a complex task serve at multiple levels: at a cognitive level they help keep the player informed and synchronized to the game status, while at a psychological level they reward, motivate and challenge. Paraskevopoulos et al [39] suggest that encouraging and motivating feedback increases game appeal and decreases the risk of patients abandoning the game and the benefits reaped from practicing. Corroborating evidence from [28] emphasizes the importance of visual and auditory feedback as an effective mechanism to identify game progress and help a player become more effective. In fact, patients that could not complete tasks successfully felt frustrated and unhappy, suggesting that players should be rewarded also for their effort and be encouraged to keep trying. It follows that exergames developed for PD patients should avoid any kind of negative feedback [31].

### 4.6. Guidelines and Navigation

Home-based exergames must provide clear and sufficient instructions to allow for unguided patient participation. Instructions can be in the form of videos or text. As an example, Assad et al [28] created build-in instructions in the form of video tutorials for each game they developed. Instruction must also be included in any interaction of the patient with the system outside the gameplay itself. Interfaces such as quit or replay require specific visual instructions, guiding players into performing the correct action. Menu navigation must be equally simple and intuitive, helping players quickly understand what they have to do to navigate the game. Assad et al report that some participants inadvertently selected unwanted menu items due to the tremor in their upper limbs, suggesting that hand hovering over a specific region of the screen to select a menu item must be avoided. Navigation-related movements must be carefully selected to avoid accidental selections due to involuntary motion of the upper limbs.

## 5. CONCLUSIONS

Physiotherapy is very important for controlling the symptoms of Parkinson's disease (PD), a thus far incurable condition that only deteriorates with age. It is exactly for this reason that traditional forms of physiotherapy are not only costly (as physiotherapy-related costs pile up) but also prone to fail in the medium to long term as patients lose interest in performing the same mundane tasks and exercises day after day. Indeed, patients following a long-term repetitive exercise schedule can easily get bored, lose interest and eventually drop out of a rehabilitation program (e.g., [28] and [32]). However, technology-supported physiotherapy is possible today and promises to liven up a daily exercise schedule. Today's game consoles such as Nintendo's Wii, Sony PlayStation Eye and Microsoft's Kinect sensor present new opportunities to infuse motivation and variety to an otherwise mundane physiotherapy routine. Such, "exergames" engage patients into repeatedly executing simple or complex exercise patterns within a goal-oriented enjoyable context, with real-time feedback and rewards in the form of visual and auditory cues. However designing safe, motivating exergames for PD patients with measurable rewards in controlling the symptoms of the disease is a challenging task. In this paper we have discussed findings of recent works and studies and have identified a number of good practices and design principles that provide for a safe and effective game experience for the end user.